\documentclass[5p]{elsarticle}
 \usepackage{axodraw}
 \usepackage{amssymb}
 \usepackage{graphicx}
 \usepackage{epic}
 \usepackage{verbatim}

 \newcommand{\jouref}[4]{{#1 }{\bf #2} (#3) #4}
 \newcommand\ibid[3]{{ibid.\ }{\bf #1} (#2) #3}
 \newcommand{\hepph}[1]{{hep-ph/#1}}

 \title{U(1)$_\mathrm{V}\otimes$U(1)$_\mathrm{A}$ symmetry breaking in superconductivity}

 \author{Kosuke Odagiri}

 \address{Electronics and Photonics Research Institute,
 National Institute of Advanced Industrial Science and Technology, 
 Tsukuba Central 2,
 1--1--1 Umezono, Tsukuba, Ibaraki 305--8568, Japan}

\begin{document}

 \begin{abstract}
  We argue that the general symmetry-breaking pattern in (quasi-)conventional
(parity and time-reversal symmetric single-band spin-singlet) superconductivity
is given by U(1)$_\mathrm{V}\otimes$U(1)$_\mathrm{A}\to$U(1)$_\mathrm{A}$,
where V stands for vector and A stands for axial-vector, as opposed to
the breaking of U(1)$_\mathrm{V}\equiv$U(1)$_\mathrm{ele/mag}$ by
itself as is commonly thought.
  This symmetry-breaking pattern implies that there will be a Higgs mode which,
together with the Goldstone
boson that is absorbed by the photon (Meissner effect), characterize the
symmetry-breaking dynamics.
  We obtain a number of strikingly simple analytical results, which
amalgamate the findings of the standard BCS and Ginzburg--Landau theories.
 \end{abstract}


 \maketitle

 \section{Introduction}

  We present a general analysis of the symmetry-breaking pattern found
in (quasi-)conventional superconductivity (SC) \cite{BCS,AGD}.
  By (quasi-)conventional SC, we mean that
the system can be considered to be invariant under parity (P) and time-reversal (T)
transformations and consist of only one band
of charge carriers, and that the SC pairs are spin-singlet. The pairing mechanism is
not restricted to phonon exchange. Generalization is
conceivable but will not be discussed in this study. In addition, we mainly
consider s-wave SC. However, generalization will be discussed.

  The charge carriers will be called `electrons' and their charge conjugate
will be called `holes', with the understanding that the analysis applies
equally to the reverse case.

  We argue that the symmetry-breaking pattern is given by
 \begin{equation}
  \mathrm{U}(1)_\mathrm{V}\otimes
  \mathrm{U}(1)_\mathrm{A}\to
  \mathrm{U}(1)_\mathrm{A},
  \label{eqn_symmetry_groups}
 \end{equation}
where V stands for vector and A stands for axial-vector, as opposed to
the breaking of U(1)$_\mathrm{V}\equiv$U(1)$_\mathrm{ele/mag}$ by
itself as is commonly thought\footnote{Strictly speaking, the left-hand side
of eqn.~(\ref{eqn_symmetry_groups}) should be divided by Z$_2$ because
of the existence of a non-trivial centre of the group.}. By the Goldstone
theorem, there will be one gapless excitation mode corresponding to the
broken symmetry, i.e., the Goldstone
boson, which is responsible for the Meissner effect, and one excitation
mode, corresponding to the residual symmetry, with finite gap, i.e.,
the Higgs boson, which is an essential
component of the symmetry breaking.

  This identification allows us to evaluate the symmetry-breaking
parameters analytically in an essentially non-perturbative (quasi-perturbative) way, by using
the framework that has been developed by Gribov, myself and Das
\cite{gribovewsb,odagirimagnetism,DO}.
  In this framework, the parameters are calculated in terms of the
dynamical degrees of freedom which, in the SC state, are the mixed states of
electrons and holes.

  The mixing of states and the resulting dynamical degrees of freedom
will be discussed in sec.~\ref{sec_dof}.
  The non-perturbative analysis will be carried out in secs.~\ref{sec_symmetries} onwards.
  The conclusions are stated at the end.

 \section{The fermionic degrees of freedom}
 \label{sec_dof}

  Because the SC condensate (Cooper pair) is charged, charge conservation
is violated in the electronic modes. That is, the system as a whole must conserve
charge, but an electron can transform into a hole by emitting a Cooper pair.
  We thus have the following two two-point functions:
 \begin{equation}
  \begin{picture}(180,40)(0,0)
   \ArrowLine(10,10)(40,10)
   \GCirc(40,10){3}{0}
   \ArrowLine(70,10)(40,10)
   \Text(40,15)[b]{$e^{-i\phi_\mathrm{SC}}\Delta_\mathrm{SC}/2$}
   \Text(90,15)[c]{and}
   \ArrowLine(140,10)(110,10)
   \GCirc(140,10){3}{0}
   \ArrowLine(140,10)(170,10)
   \Text(140,15)[b]{$e^{+i\phi_\mathrm{SC}}\Delta_\mathrm{SC}/2$}
  \end{picture},
  \label{eqn_two_point_mixing}
 \end{equation}
  with spin conservation in each case.
  $\phi_\mathrm{SC}$ is the phase of the SC condensate, and $\Delta_\mathrm{SC}$ is the
absolute value of the superconducting gap. We may adopt
$\phi_\mathrm{SC}=0$ without loss of generality.
  As in the case of BCS theory \cite{BCS,AGD}, the pairing
interaction may have a finite energy range, and this restricts the applicability of
eqn.~(\ref{eqn_two_point_mixing}) to be within that energy range. In this sense,
and also in the sense of non-s-wave pairing symmetries, $\Delta_\mathrm{SC}$
is, in general, $\mathbf{k}$-dependent. $\mathbf{k}$ is the wavevector. Note that
the SC condensate itself is not a dynamical degree of freedom, even though the
fluctuations in the SC condensate, i.e., the Goldstone and Higgs modes, are
dynamical.

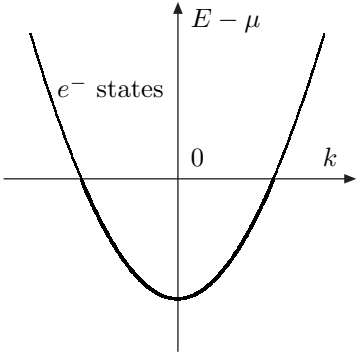
\begin{figure}[ht]
 \begin{picture}(150,150)(0,0)
  \LongArrow(75,10)(75,140)
  \LongArrow(10,75)(140,75)
  \qbezier(75,30)(100,30)(130,130)
  \qbezier(75,30)(50,30)(20,130)
%  \Text(5,140)[l]{(a)}
  \Text(80,135)[l]{$E-\mu$}
  \Text(135,80)[rb]{$k$}
  \Text(80,80)[bl]{0}
  \Text(30,110)[l]{$e^-$ states}
  \thicklines
  \qbezier(75,30)(92,30)(111,75)
  \qbezier(75,30)(58,30)(39,75)
 \end{picture}
 \caption{\label{fig_dispersion_a}The dispersion relation in an ordinary metal (schematic). The
zone boundaries are not shown. Thick line indicates occupied states.}
 \end{figure}

  Let us apply eqn.~(\ref{eqn_two_point_mixing}) to
the mixing of states using the Nambu representation \cite{namburepresentation}.
  We start from the case of ordinary metals (fig.~\ref{fig_dispersion_a}).
  The dispersion relation can be denoted as:
 \begin{equation}
  \left(\psi^*_\uparrow \ \psi^*_\downarrow \right)\left(
  \begin{array}{cc} \xi(\mathbf{k}) & 0 \\
  0 & \xi(\mathbf{k}) \end{array} \right)
  \left(\begin{array}c \psi_\uparrow \\ \psi_\downarrow
  \end{array}\right).
  \label{eqn_ordinary_dispersion_relation}
 \end{equation}
  Note that $\xi(\mathbf{k})\equiv \epsilon(\mathbf{k})-\mu$.

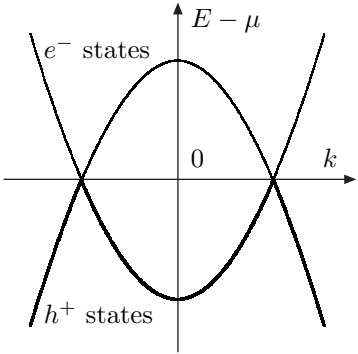
\begin{figure}[ht]
 \begin{picture}(150,150)(0,0)
  \LongArrow(75,10)(75,140)
  \LongArrow(10,75)(140,75)
  \qbezier(75,30)(100,30)(130,130)
  \qbezier(75,120)(100,120)(130,20)
  \qbezier(75,30)(50,30)(20,130)
  \qbezier(75,120)(50,120)(20,20)
%  \Text(5,140)[l]{(b)}
  \Text(80,135)[l]{$E-\mu$}
  \Text(135,80)[rb]{$k$}
  \Text(80,80)[bl]{0}
  \Text(25,125)[l]{$e^-$ states}
  \Text(25,25)[l]{$h^+$ states}
  \thicklines
  \qbezier(75,30)(92,30)(111,75)
  \qbezier(75,30)(58,30)(39,75)
  \qbezier(111,75)(121,50)(130,20)
  \qbezier(39,75)(29,50)(20,20)
 \end{picture}
 \caption{\label{fig_dispersion_b}The half-inverted dispersion relation.}
 \end{figure}

  We now take one of the two spin states, let us say $\psi_\downarrow$, and
make a (CPT, where C stands for charge conjugation) conjugate of this state, such that:
 \begin{equation}
  \psi_\downarrow\longrightarrow\psi^*_\downarrow
  \equiv(\psi_\downarrow)^\mathrm{CPT}
  \equiv(\psi_\downarrow)^*\equiv(\psi^*)_\uparrow.
 \end{equation}
  That is, instead of considering the presence of a $\psi_\downarrow$ state, we
consider the absence of a hole state, $(\psi^*)_\uparrow$, which annihilates the
$\psi_\downarrow$ state.  The dispersion relation of eqn.~(\ref{eqn_ordinary_dispersion_relation})
is now replaced by the half-inverted dispersion relation:
 \begin{equation}
  \left(\psi^*_\uparrow \ \psi_\downarrow \right)\left(
  \begin{array}{cc} \xi(\mathbf{k}) & 0 \\
  0 & -\xi(-\mathbf{k}) \end{array} \right)
  \left(\begin{array}c \psi_\uparrow \\ \psi^*_\downarrow
  \end{array}\right).
  \label{eqn_half_inverted_dispersion_relation}
 \end{equation}
  This is depicted in fig.~\ref{fig_dispersion_b}.

  In the SC state, the presence of the two-point functions,
as given by eqn.~(\ref{eqn_two_point_mixing}), implies
that there will be off-diagonal terms in eqn.~(\ref{eqn_half_inverted_dispersion_relation}),
that are given by
 \begin{equation}
  \left(\psi^*_\uparrow \ \psi_\downarrow \right)\left(
  \begin{array}{cc} \xi(\mathbf{k}) & \Delta_\mathrm{SC}/2 \\
  \Delta_\mathrm{SC}/2 & -\xi(-\mathbf{k}) \end{array} \right)
  \left(\begin{array}c \psi_\uparrow \\ \psi^*_\downarrow
  \end{array}\right).
  \label{eqn_superconducting_dispersion_relation}
 \end{equation}
  We have adopted $\phi_\mathrm{SC}=0$

\begin{figure}[ht]
 \begin{picture}(150,150)(0,0)
  \LongArrow(75,10)(75,140)
  \LongArrow(10,75)(140,75)
  \qbezier(75,120)(87,120)(100,97)
  \qbezier(100,97)(106.5,85.5)(110,85.5)
  \qbezier(110,85.5)(113.5,85.5)(130,130)
  \qbezier(75,120)(63,120)(50,97)
  \qbezier(50,97)(43.5,85.5)(40,85.5)
  \qbezier(40,85.5)(36.5,85.5)(20,130)
  \LongArrow(40,75)(40,83)
  \LongArrow(40,75)(40,67)
  \thicklines
  \qbezier(75,30)(87,30)(100,53)
  \qbezier(100,53)(106.5,64.5)(110,64.5)
  \qbezier(110,64.5)(113.5,64.5)(130,20)
  \qbezier(75,30)(63,30)(50,53)
  \qbezier(50,53)(43.5,64.5)(40,64.5)
  \qbezier(40,64.5)(36.5,64.5)(20,20)
%  \Text(5,140)[l]{(c)}
  \Text(80,135)[l]{$E-\mu$}
  \Text(135,80)[rb]{$k$}
  \Text(80,80)[bl]{0}
  \Text(25,125)[l]{mixed states}
  \Text(25,25)[l]{mixed states}
  \Text(44,72)[lt]{$\Delta_\mathrm{SC}$}
 \end{picture}
 \caption{\label{fig_dispersion_c}The dispersion relation in the
SC state.
 }
 \end{figure}
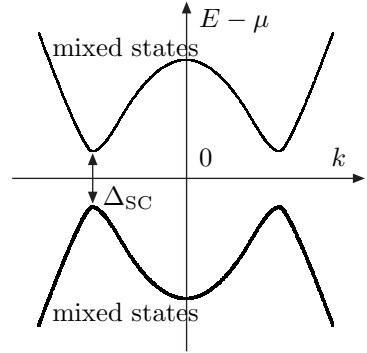

  Let us make use of the parity invariance of the system, which implies
$\xi(-\mathbf{k})=\xi(\mathbf{k})$. We define the physical states
$\psi_\mathrm{u,d}$ and mixing
angle $\theta_\mathrm{SC}(\mathbf{k})$ (abbreviated as $\theta$ in the
following) by
 \begin{equation}
  \left(\begin{array}c\psi_\mathrm{u}\\\psi_\mathrm{d}\end{array}\right)=
  \left(\begin{array}{cc}\cos\theta & \sin\theta\\ -\sin\theta & \cos\theta
  \end{array}\right)\left(\begin{array}c\psi_\uparrow\\\psi^*_\downarrow
  \end{array}\right).
  \label{eqn_mixing_angle_definition}
 \end{equation}
  The diagonalization of eqn.~(\ref{eqn_half_inverted_dispersion_relation})
then yields
 \begin{equation}
  \left(\psi^*_\mathrm{u} \ \psi^*_\mathrm{d} \right)\left(
  \begin{array}{cc} \widetilde\xi(\mathbf{k}) & 0 \\
  0 & -\widetilde\xi(\mathbf{k}) \end{array} \right)
  \left(\begin{array}c \psi_\mathrm{u} \\ \psi_\mathrm{d}
  \end{array}\right),
  \label{eqn_diagonalized_dispersion_relation}
 \end{equation}
  where
 \begin{equation}
  \widetilde\xi(\mathbf{k})=
  \sqrt{\xi(\mathbf{k})^2+\left(\frac{\Delta_\mathrm{SC}}2\right)^2}
 \end{equation}
  and the mixing angle is given by:
 \begin{equation}
  \tan2\theta_\mathrm{SC}(\mathbf{k})=\frac{\Delta_\mathrm{SC}}{2\xi(\mathbf{k})}.
 \end{equation}
  This gives rise to the structure that is shown in fig.~\ref{fig_dispersion_c}.
%  Let us adopt the following convention for $\theta_\mathrm{SC}$:
% \begin{equation}
%  2\theta_\mathrm{SC}(\mathbf{k})=
%  \pi/2-\tan^{-1}\frac{2\xi(\mathbf{k})}{\Delta_\mathrm{SC}},
% \end{equation}
%  so that $\theta\to\pi/2$ as $\xi\to-\infty$ and $0$ as $\xi\to0$.

  Note that the gap is given by $\Delta_\mathrm{SC}$, as it should.
%  However, we notice that the excitations are not of the form $e^-e^-$, as
%would be expected naively. The excitations will be a mixture of $e^-e^-$,
%$h^+h^+$ and $e^-h^+$, with the average charge of excitation
%being equal to zero. This should be tested if possible, either experimentally
%or numerically.

 \section{Symmetries of the system}
 \label{sec_symmetries}

  Let us now focus on symmetries. Note that the usual electromagnetic
U(1) symmetry, the U(1)$_\mathrm{V}$ symmetry that is, is now
associated with the rotation
 \begin{equation}
  \left(\begin{array}c \psi_\uparrow \\ \psi^*_\downarrow
  \end{array}\right)\to
  \exp\left(i\alpha_\mathrm{V}\sigma_3\right)
  \left(\begin{array}c \psi_\uparrow \\ \psi^*_\downarrow
  \end{array}\right).
 \end{equation}
  The presence of mixing implies that this is no longer a conserved symmetry.
  On the other hand, we can define another U(1) symmetry, i.e., the
U(1)$_\mathrm{A}$ symmetry, which is associated with the rotation
 \begin{equation}
  \left(\begin{array}c \psi_\uparrow \\ \psi^*_\downarrow
  \end{array}\right)\to
  \exp\left(i\alpha_\mathrm{A}\right)
  \left(\begin{array}c \psi_\uparrow \\ \psi^*_\downarrow
  \end{array}\right).
 \end{equation}
  This is conserved even with the mixing of the two states, because both states
transform in the same way.

  The U(1)$_\mathrm{A}$ group is an integral component of the symmetry-breaking
pattern. To demonstrate this, consider the Goldstone boson that is associated
with the breaking of the U(1)$_\mathrm{V}$ symmetry. We shall demonstrate
later that its coupling to fermions is proportional to $\sigma_2$. However,
$\sigma_2$ is not invariant under U(1)$_\mathrm{V}$ rotation, and 
a linear combination of $\sigma_1$ and $\sigma_2$ arises, in general, under
this rotation. There are therefore (at least) two bosonic modes that couple to fermions and,
if one linear combination of $\sigma_1$ and $\sigma_2$ corresponds to the
Goldstone boson, there must be the other linear combination, which corresponds to
a mode with finite excitation energy, i.e., the
Higgs boson. The existence of a non-elementary Higgs boson requires that
there is a corresponding
conserved symmetry which, in this case, is U(1)$_\mathrm{A}$.
 In short, a closedness condition of the symmetry group implies that the
symmetry is given by U(1)$_\mathrm{V}\otimes$U(1)$_\mathrm{A}$.

  The U(1)$_\mathrm{A}$ transformation resembles the SU(2)$_\mathrm{spin}$ rotation
around the $z$ axis, but is not the same. U(1)$_\mathrm{A}$ rotation is the
phase rotation that acts in the opposite direction after P conjugation. The above
formulation, which adopts the Nambu representation, is merely a means.
The same rotation can be defined by, for instance,
dividing the phase space into two, and applying opposite phase rotation to each
of the two sections.

\begin{figure}[ht]
 \begin{picture}(150,150)(0,0)
  \LongArrow(75,10)(75,140)
  \LongArrow(10,75)(140,75)
  \qbezier(75,20)(100,20)(130,120)
  \qbezier(75,20)(50,20)(20,120)
  \qbezier(75,110)(50,110)(20,10)
  \qbezier(75,110)(100,110)(130,10)
  \DashLine(25,65)(55,65){4}
  \LongArrow(30,82)(30,76)
  \Line(30,75)(30,65)
  \LongArrow(30,58)(30,64)
  \Text(25,73)[tr]{$\frac{\Delta_\mathrm{ex}}2$}
%  \Text(5,140)[l]{(d)}
  \Text(80,135)[l]{$E-\mu$}
  \Text(135,80)[rb]{$k$}
  \Text(80,80)[bl]{0}
  \Text(24,120)[l]{$e^-$ states}
  \Text(25,15)[l]{$h^+$ states}
  \thicklines
  \qbezier(75,20)(94,20)(115,75)
  \qbezier(75,20)(56,20)(35,75)
  \qbezier(106,75)(122,40)(130,10)
  \qbezier(44,75)(28,40)(20,10)
 \end{picture}
 \caption{\label{fig_dispersion_d}The dispersion relation for (ferro-)magnetic metals.}
 \end{figure}
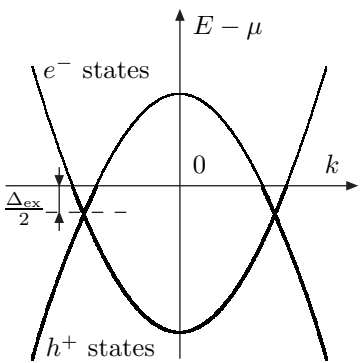

  Having said that, the above formulation makes it easy to make comparisons with
magnetism in metals, especially ferromagnetism. Magnetism is where the energy
level of a state is split according to spin by an exchange energy $\Delta_\mathrm{ex}$.
In our formulation, this is described by the dispersion relation
 \begin{equation}
  \left(\psi^*_\uparrow \ \psi_\downarrow \right)\left(
  \begin{array}{cc} \xi(\mathbf{k})-\frac{\Delta_\mathrm{ex}}2 & 0 \\
  0 & -\xi(-\mathbf{k})-\frac{\Delta_\mathrm{ex}}2 \end{array} \right)
  \left(\begin{array}c \psi_\uparrow \\ \psi^*_\downarrow
  \end{array}\right).
  \label{eqn_magnetic_dispersion_relation}
 \end{equation}
  This is shown in fig.~\ref{fig_dispersion_d}

  In the case of magnetism, the broken symmetry leads to the
non-conservation of the spin current, which is restored by including
the contribution of the Goldstone magnons. The same should be the
case in superconductivity, where the vector U(1) current becomes non-conserved.

  Let us consider this in more detail. In the $(\psi,\psi^*)$ basis, the
non-conservation of current arises from the off-diagonal sector, which
must be cancelled by the Goldstone-boson contribution. We first define
the form factor $f\equiv f_\mathrm{SC}$ as the coupling strength
of the current--Goldstone-boson two-point function where, in the
four-component notation,
 \begin{equation}
  \begin{picture}(120,15)(0,0)
   \Text(5,5)[l]{$k_\mu\times$}
%   \Text(2,0)[tl]{$\longrightarrow$}
   \GCirc(30,5){3}{0}
   \Text(30,9)[b]{$\mu$}
   \DashLine(30,5)(60,5){5}
   \Text(65,5)[l]{$=ifD_G(k).$}
   \label{eqn_form_factor_defintion}
  \end{picture}
 \end{equation}
  $k$ flows left to right. $D_G$ is the Goldstone-boson propagator.
$f$ and $D_G$ can be calculated by evaluating the two-point function,
as we shall demonstrate later.
  Note that the Goldstone boson will be absorbed by the photon, which process
being the cause of the Meissner effect.

  The non-conservation of current and its restoration can be seen in various ways.
One possibility, which we consider to be the simplest, is to consider
the following Ward--Takahashi identity (in the $(\psi,\psi^*)$ basis):
 \begin{equation}
  \begin{picture}(205,30)(0,0)
   \Text(5,15)[l]{$k_\mu\Biggl[$}
   \ArrowLine(30,5)(50,5)
   \ArrowLine(70,5)(50,5)
   \DashLine(50,5)(50,20){5}
   \GCirc(50,20){3}{0}
   \Text(50,25)[b]{$\mu$}
   \Text(75,15)[c]{$+$}
   \ArrowLine(80,5)(90,18)
   \ArrowLine(90,18)(105,18)
   \ArrowLine(120,18)(105,18)
   \GCirc(105,18){3}{0}
   \Text(87,19)[b]{$\Gamma^\mu$}
   \Text(105,14)[t]{$\frac{\Delta_\mathrm{SC}}2$}
   \Text(127.5,15)[c]{$+$}
   \ArrowLine(135,18)(150,18)
   \ArrowLine(165,18)(150,18)
   \ArrowLine(175,5)(165,18)
   \GCirc(150,18){3}{0}
   \Text(170,19)[b]{$\Gamma^\mu$}
   \Text(150,14)[t]{$\frac{\Delta_\mathrm{SC}}2$}
   \Text(180,15)[l]{$\Biggr]=0.$}
  \end{picture}
  \label{eqn_mixing_ward_identity}
 \end{equation}
  $\Gamma^\mu$ refers to the vector current vertex ($\Gamma^0=1$), which by
itself must satisfy the Ward--Takahashi identity, as usual.
  However, the sum of the second and the third terms does not
satisfy the Ward--Takahashi identity, and the residual contribution must
be cancelled by the first term. The same applies to the set of amplitudes
in which the directions of the arrows are reversed. These two conditions fix the coupling of
the Goldstone boson to fermions, which is found to have the
form
 \begin{equation}
  f^{-1}\Delta_\mathrm{SC}\left(\psi^*_\uparrow \ \psi_\downarrow \right)\left(
  \begin{array}{cc} 0 & -i \\ i & 0 \end{array} \right)
  \left(\begin{array}c \psi_\uparrow \\ \psi^*_\downarrow
  \end{array}\right)G.
  \label{eqn_Goldstone_fermion_coupling}
 \end{equation}
  The coupling in the $(\psi_u,\psi_d)$ basis can be found by
rotation. In particular, in the soft limit of the Goldstone boson, the
coupling is completely off-diagonal. That is, $\sigma_2$ is proportional
to, and commutes with, the generator of the $\theta_\mathrm{SC}$ rotation.

  The non-vanishing coupling of the Goldstone boson in the
soft limit, which apparently is in contradiction with the so-called
decoupling theorem, should not worry us, as the Ward--Takahashi
identities take care of such pathological contributions. However, we
need to introduce the Higgs degree of freedom to guarantee this.
  As explained before, this Higgs boson is associated with the
residual U(1)$_\mathrm{A}$ symmetry.

  The coupling of the Higgs to the fermions must have the same
strength as the Goldstone boson, and must be of the same form as
the $\Delta_\mathrm{SC}$ coupling. That is, the coupling is given by
 \begin{equation}
  f^{-1}\Delta_\mathrm{SC}\left(\psi^*_\uparrow \ \psi_\downarrow \right)\left(
  \begin{array}{cc} 0 & 1 \\ 1 & 0 \end{array} \right)
  \left(\begin{array}c \psi_\uparrow \\ \psi^*_\downarrow
  \end{array}\right)(v+H).
  \label{eqn_Higgs_fermion_coupling}
 \end{equation}
  One can use Ward--Takahashi identities to derive this more rigorously.
  Here, $v$ is a constant, and carries the meaning of the vacuum
expectation value of the SC condensate. $f=2v$ in order that this
equation is consistent with eqn.~(\ref{eqn_two_point_mixing}).
  It follows that the vacuum expectation value of the SC condensate,
and the penetration depth, can be calculated by finding the
current--Goldstone-boson two-point function.

 \section{Two-point functions}

  There being only one Goldstone degree of freedom implies that
the Goldstone boson must have a linear dispersion relation, for small
$\mathbf{k}$. Let us parametrize $D_G$ as:
 \begin{equation}
  D_G(E,\mathbf{k})=\frac1{E^2-u^2\mathbf{k}^2+i0}.
  \label{eqn_goldstone_propagator}
 \end{equation}
  For this form of the Goldstone-boson Green's function, the
Higgs-boson Green's function can be assumed to be given by:
 \begin{equation}
  D_H(E,\mathbf{k})=\frac1{E^2-u^2\mathbf{k}^2-\Delta_H^2+i0}.
  \label{eqn_higgs_propagator}
 \end{equation}
  This is probably only an approximation, but this approximation
makes it easy to impose current conservation.
  Specifically, we would like the following vertex to conserve the
Ward--Takahashi identity:
 \begin{equation}
  %ins \includegraphics{diagram13.eps}
  \begin{picture}(210,50)(0,0)
   \DashLine(5,25)(30,25){5}
   \Text(7,29)[b]{$G$}
   \Text(30,21.5)[c]{\huge *} \Text(30,32)[b]{$\mu$}
   \DashLine(30,25)(55,25){5}
   \Text(53,29)[b]{$H$}
   \Text(65,27)[c]{$=$}
   \DashLine(80,25)(105,25){5}
   \Text(82,29)[b]{$G$}
   \GCirc(105,25){3}{0}
   \Text(105,32)[b]{$\mu$}
   \Text(105,20)[t]{$-2i(k_G+k_H)^\mu$}
   \DashLine(105,25)(130,25){5}
   \Text(128,29)[b]{$H$}
   \Text(145,27)[c]{$+$}
   \Text(155,8)[b]{$G$}
   \DashLine(155,5)(180,15){5}
   \DashLine(180,15)(205,5){5}
   \Text(205,8)[b]{$H$}
   \DashLine(180,15)(180,35){5}
   \Text(177,23)[r]{$G$}
   \GCirc(180,35){3}{0}
   \Text(180,42)[b]{$\mu$}
  \end{picture}.
  \label{eqn_modified_vertex_bosonic}
 \end{equation}
  The normalization of the vertex factor $-2i(k_G+k_H)^\mu$ ($u=1$ for simplicity;
all momenta flowing left to right) is determined by the requirement of the following
Ward--Takahashi identity, again in ($\psi,\psi^*$) basis, which
is a straightforward modification of eqn.~(\ref{eqn_mixing_ward_identity}):
 \begin{equation}
  \begin{picture}(205,40)(0,0)
   \Text(5,20)[l]{$k_\mu\Biggl[$}
   \ArrowLine(30,5)(50,5)
   \ArrowLine(70,5)(50,5)
   \DashLine(50,5)(50,35){5}
   \Text(50,17.5)[c]{\huge *}
%   \GCirc(50,20){3}{0}
   \Text(46,20)[r]{$\mu$}
   \Text(53,13)[l]{$G$}
   \Text(53,30)[l]{$H$}
   \Text(75,15)[c]{$+$}
   \ArrowLine(80,5)(90,18)
   \ArrowLine(90,18)(105,18)
   \ArrowLine(120,18)(105,18)
   \DashLine(105,18)(105,35){5}
   \Text(108,30)[l]{$H$}
   \Text(87,19)[b]{$\Gamma^\mu$}
   \Text(127.5,15)[c]{$+$}
   \ArrowLine(135,18)(150,18)
   \ArrowLine(165,18)(150,18)
   \ArrowLine(175,5)(165,18)
   \DashLine(150,18)(150,35){5}
   \Text(148,30)[r]{$H$}
   \Text(170,19)[b]{$\Gamma^\mu$}
   \Text(180,15)[l]{$\Biggr]=0.$}
  \end{picture}
  \label{eqn_higgs_ward_identity}
 \end{equation}
  Note that the form of the current is consistent with the form expected
for the kinetic-energy term of the effective Lagrangian density:
 \begin{equation}
  \mathcal{L}_\mathrm{kin}=
  \frac14\left|\left(i\hbar\frac{\partial}{\partial t}-
  2eA_0\sigma_3\right)\Phi\right|^2
  -
  \frac{u^2}{4}\left|\left(-i\hbar\nabla-
  2e\mathbf{A}\sigma_3\right)\Phi\right|^2,
  \label{eqn_kinetic_term}
 \end{equation}
% \begin{eqnarray}
%  \mathcal{L}_\mathrm{kin}&=&
%  \frac14\left|\left(i\hbar\frac{\partial}{\partial t}-
%  2eA_0\sigma_3\right)\Phi\right|^2
%  \nonumber\\&-&
%  \frac{u^2}{4}\left|\left(-i\hbar\nabla-
%  2e\mathbf{A}\sigma_3\right)\Phi\right|^2,
%  \label{eqn_kinetic_term}
% \end{eqnarray}
  where $\Phi=(v+H)\sigma_1+G\sigma_2$.

  Using the Ward--Takahashi identity associated with
eqn.~(\ref{eqn_modified_vertex_bosonic}), we then find
that the $GGH$ three-point coupling is given by:
 \begin{equation}
  g_{GGH}=2f^{-1}\Delta_H^2=\Delta_H^2/v.
 \end{equation}
  Similarly to ref.~\cite{DO}, we find that this is consistent with
the following form of the multi-bosonic Lagrangian density:
 \begin{equation}
  \mathcal{L}_\mathrm{bosons}=-
  \frac{\Delta_H^2}{8v^2}(G^2+(H+v)^2-v^2)^2.
  \label{eqn_multiboson_lagrangian}
 \end{equation}
  We shall not derive the other terms comprising this equation
explicitly, since doing so will be merely repeating ref.~\cite{DO}.
%  However, it seems to the author that symmetry by itself
%is sufficient to complete eqn.~(\ref{eqn_multiboson_lagrangian}).

  As for the current--Goldstone-boson two-point function, a
smarter method than to calculate this blindly is to calculate the current--current
two-point function \cite{gribovlongshort}.
  We shall omit the intermediate steps (see ref.~\cite{odagirimagnetism}),
but this gives us the following (one-loop) relation:
 \begin{equation}
  f^2=-2\int\frac{d^{d+1}k}{(2\pi)^{d+1}i}(\sin2\theta_\mathrm{SC}(\mathbf{k}))^2
  G_\mathrm{u}(k)G_\mathrm{d}(k),
 \end{equation}
  which is valid for small energy and momenta.
  We have neglected the contribution due to bosonic loops. We expect
the bosonic loops to be suppressed because the Goldstone boson is absorbed by
the photon.
  Note that the u states are empty and d states are occupied, so that:
 \begin{equation}
  f^2=2\int\frac{d^d\mathbf{k}(\sin2\theta_\mathrm{SC}(\mathbf{k}))^2}
  {(2\pi)^d2\sqrt{(\Delta_\mathrm{SC}/2)^2+\xi^2)}}.
  \label{eqn_form_factor_initial}
 \end{equation}
  Let us replace the phase-space integration by:
 \begin{equation}
  2\frac{d^d\mathbf{k}}{(2\pi)^d}=g(\mu+\xi)d\xi\approx g_Fd\xi.
 \end{equation}
  Here, $g(\mu+\xi)$ refers to the density of states before
symmetry breaking.
  We then obtain
 \begin{equation}
  f^2\approx\frac{g_F}2
  \int_{-\infty}^{\infty}d(2\xi/\Delta_\mathrm{SC})
  \left[1+(2\xi/\Delta_\mathrm{SC})^2\right]^{-3/2}.
 \end{equation}
  Strictly speaking, the limits of integration should be given by the relevant energy scale,
e.g.\ the Debye frequency in the case of BCS superconductors.
However, the integral is convergent and therefore we can take the
limits to be $\pm\infty$. We then obtain
 \begin{equation}
  f^2=4v^2\approx g_F.
  \label{eqn_vev_normal_state_dos}
 \end{equation}
  That is, to the first approximation, the vacuum expectation value of the
SC condensate is given solely by the electronic density of
states in the normal state. This surprising conclusion implies that the
penetration depth $\lambda$, for $d=3$, is given by
 \begin{equation}
  \lambda=1/\sqrt{\mu_0u^2e^2g_F}.
  \label{eqn_penetration_depth}
 \end{equation}
  This conclusion, as well as eqn.~(\ref{eqn_vev_normal_state_dos}),
is independent of the pairing symmetry. That is, $\Delta_\mathrm{SC}$
in the above equations can have an angular dependence without altering
the results.

  The effective Lagrangian terms for contributions other than the
fermionic contributions (which also can be written down but decouple
in the SC phase) are given by
 \begin{equation}
  \mathcal{L}=\mathcal{L}_\mathrm{bosons}+\mathcal{L}_\mathrm{ele/mag}
  +\mathcal{L}_\mathrm{kin}.
 \end{equation}
  The first term is given by eqn.~(\ref{eqn_multiboson_lagrangian}). The
second term is the usual Maxwellian term. The third term is given by
eqn.~(\ref{eqn_kinetic_term}).

  By comparing against the corresponding Ginzburg--Landau terms,
we then obtain eqn.~(\ref{eqn_penetration_depth}) for the penetration
depth, and
 \begin{equation}
  \xi=\frac{u\hbar\sqrt2}{\Delta_H},
  \label{eqn_coherence_length}
 \end{equation}
  for the coherence length.
  The Ginzburg--Landau parameter $\kappa$ is given by
 \begin{equation}
  \kappa=\lambda/\xi=\Delta_H/u^2\hbar\sqrt{2\mu_0e^2g_F}.
  \label{eqn_gl_kappa}
 \end{equation}

 \section{Tadpole cancellation and Higgs excitation energy}

  We now consider the tadpole cancellation condition:
 \begin{equation}
  %ins \includegraphics{}
  \begin{picture}(155,32)(0,0)
   \DashLine(5,15)(30,15){5}
   \ArrowArcn(45,15)(15,270,-90)
   \Text(70,15)[c]{$+$}
   \DashLine(80,15)(105,15){5}
   \DashCArc(120,15)(15,0,360){5}
   \Text(143,15)[l]{$=0.$}
  \end{picture}
  \label{eqn_tadpole_cancellation}
 \end{equation}
  The cancellation of the tadpole is a vacuum-stability condition.
  The second loop contains both Goldstone and Higgs bosons.
  This particular cancellation strategy is not the only possibility.
Other mechanisms may be possible,
depending on the nature of the pairing interaction. However,
eqn.~(\ref{eqn_tadpole_cancellation}) should be understood as
an all-order condition. This implies that the first loop includes
all self-energy type corrections. Hence the only other possibility
for cancelling the tadpole is, in our opinion, the mixing
of the Higgs boson with some other bosonic mode that has the
same quantum numbers as the vacuum (see ref.~\cite{odagirimagnetism}).

  The fermionic loop of eqn.~(\ref{eqn_tadpole_cancellation})
yields
 \begin{equation}
  -\int\frac{d^{d+1}k}{(2\pi)^{d+1}i}f^{-1}\Delta_\mathrm{SC}
  (-\sin2\theta_\mathrm{SC})G_\mathrm{d}(k).
 \end{equation}
  This reduces to
 \begin{equation}
  f^{-1}g_F\int d\xi\frac{(\Delta_\mathrm{SC}/2)^2}
 {\sqrt{(\Delta_\mathrm{SC}/2)^2+\xi^2}}.
  \label{eqn_tadpole_fermionic_intermediate}
 \end{equation}
  This integral is divergent, and therefore needs to be cut-off.
Cutting off the integral at $\omega_\mathrm{cut}$, we obtain
 \begin{equation}
  \frac{g_F\Delta_\mathrm{SC}^2}{2f}
  \sinh^{-1}\frac{2\omega_\mathrm{cut}}{\Delta_\mathrm{SC}}
  \approx\frac{g_F\Delta_\mathrm{SC}^2}{2f}
  \ln\frac{\omega_\mathrm{cut}}{\Delta_\mathrm{SC}}.
  \label{eqn_tadpole_fermionic}
 \end{equation}

  As for the bosonic loop, let us say that the Goldstone-boson loop
is suppressed because of the absorption of the Goldstone boson in
the photon. For the Higgs-boson contribution, we obtain:
 \begin{equation}
  \frac12\int\frac{d^{d+1}k}{(2\pi)^{d+1}i}6f^{-1}\Delta_H^2
  D_H(k).
  \label{eqn_tadpole_bosonic_initial}
 \end{equation}
  Using eqn.~(\ref{eqn_higgs_propagator}), we then obtain
 \begin{equation}
  \int\frac{-(3f^{-1}\Delta_H^2)d^d\mathbf{k}}
  {(2\pi)^d2\sqrt{u^2\mathbf{k}^2+\Delta_H^2}}.
  \label{eqn_tadpole_bosonic}
 \end{equation}
  It is possible to show, by considering the spatial component of the
vector current--current two-point function, that $u=\mathcal{O}(v_F)$,
where $v_F$ is the Fermi velocity in the normal state. Therefore
$u^2\mathbf{k}^2\gg\Delta_H^2$ near the zone boundaries where
$\left|\mathbf{k}\right|\approx K\approx\pi/a$. In
this limit, the integral is evaluated to be
 \begin{equation}
  \left\{\begin{array}{ll}
   -3\Delta_H^2K/4\pi uf & (d=2), \\
   -3\Delta_H^2K^2/8\pi^2uf & (d=3).
  \end{array}\right.
 \end{equation}

  Next, let us calculate the Higgs-boson excitation energy.
  The relevant Feynman graphs are shown in fig.~\ref{fig_higgs_self_energy}.

 \begin{figure}[ht]
  %ins \includegraphics{diagram35.eps}
  \begin{picture}(120,60)(0,0)
   \Text(60,58)[t]{(a)}
   \DashLine(20,25)(50,25){5}
   \ArrowArcn(60,25)(10,0,180)
   \ArrowArcn(60,25)(10,180,360)
   \DashLine(70,25)(100,25){5}
  \end{picture}
  %ins \includegraphics{diagram36.eps}
  \begin{picture}(120,60)(0,0)
   \Text(60,58)[t]{(b)}
   \DashLine(20,25)(50,25){5}
   \DashCArc(60,25)(10,0,360){5}
   \DashLine(70,25)(100,25){5}
  \end{picture}

  %ins \includegraphics{diagram37.eps}
  \begin{picture}(120,60)(0,0)
   \Text(60,58)[t]{(c)}
   \DashLine(20,15)(100,15){5}
   \DashCArc(60,25)(10,0,360){5}
  \end{picture}
  \caption{\label{fig_higgs_self_energy}
  The diagrams for the self-energy of the Higgs (and Goldstone) boson. 
Diagram b involves $GGH$ and $HHH$ vertexes, and
diagram c involves $GGHH$ and $HHHH$ vertexes.}
 \end{figure}

  Let us calculate $\Delta_H$ as the sum of the self-energy
contributions at zero external momentum. For the contribution
of fig.~\ref{fig_higgs_self_energy}a, we obtain
 \begin{equation}
  2\int\frac{d^{d+1}k}{(2\pi)^{d+1}i}
  (f^{-1}\Delta_\mathrm{SC}\cos2\theta_\mathrm{SC})^2
  G_\mathrm{u}(k)G_\mathrm{d}(k).
 \end{equation}
  This reduces to
 \begin{equation}
  -\frac{g_F}2\int d\xi
  \frac{(f^{-1}\Delta_\mathrm{SC}\cos2\theta_\mathrm{SC})^2}
  {\sqrt{(\Delta_\mathrm{SC}/2)^2+\xi^2}}.
 \end{equation}
  This is negative and logarithmically divergent.

  Here we find an analogy to our previous analyses \cite{odagirimagnetism,DO}
in that such divergent contributions
are cancelled by tadpoles. Corresponding to fig.~\ref{fig_higgs_self_energy}c,
we have the contribution
 \begin{equation}
  -\frac12\int\frac{d^{d+1}k}{(2\pi)^{d+1}i}12f^{-2}\Delta_H^2
  D_H(k).
  \label{eqn_higgs_selfenergy_tadpole}
 \end{equation}
  This is $-2f^{-1}$ times the bosonic tadpole that is given by
eqn.~(\ref{eqn_tadpole_bosonic_initial}). This implies that
this contribution is equal to $+2f^{-1}$ times the fermionic
tadpole contribution that is given by eqn.~(\ref{eqn_tadpole_fermionic_intermediate}).
  As a result, the net contribution to the Higgs-boson self-energy is given by
 \begin{equation}
  \frac{g_F}2\int d\xi
  \frac{(f^{-1}\Delta_\mathrm{SC}\sin2\theta_\mathrm{SC})^2}
  {\sqrt{(\Delta_\mathrm{SC}/2)^2+\xi^2}}.
 \end{equation}
  This is positive and finite, and is in fact equal to
eqn.~(\ref{eqn_form_factor_initial}) times $f^{-2}\Delta_\mathrm{SC}^2$.
  This implies that
 \begin{equation}
  \Delta_H=\Delta_\mathrm{SC}.
  \label{eqn_gap_higgs_identity}
 \end{equation}
  As for the contribution of fig.~\ref{fig_higgs_self_energy}b,
this cannot be large.
  We may write down the amplitude as
 \begin{equation}
  -\frac12\int\frac{d^{d+1}k}{(2\pi)^{d+1}i}(6f^{-2}\Delta_H^2)^2
  (D_H(k))^2,
 \end{equation}
  but using
 \begin{equation}
  (D_H(k))^2=\frac{\partial}{\partial (\Delta_H^2)} D_H(k),
 \end{equation}
  we find that the integral vanishes in the limit $uK\gg\Delta_H$.
  Having said that, eqn.~(\ref{eqn_gap_higgs_identity}) is not
protected by any symmetry, and so it is possible that this degeneracy
of $\Delta_H$ and $\Delta_\mathrm{SC}$ is lifted by corrections
particularly near the SC gap threshold.

  Since eqns.~(\ref{eqn_tadpole_fermionic}) and (\ref{eqn_tadpole_bosonic})
must cancel when added together, we now obtain
 \begin{equation}
  \Delta_{\mathrm{SC}}\approx\omega_\mathrm{cut}
  \exp\left(-\frac{3}{ug_F}
  \left(\frac{K}{2\pi}\right)^{d-1}\right).
 \end{equation}
  This is identical to the BCS result \cite{AGD}, but with the phonon
coupling $\Lambda$ replaced by
 \begin{equation}
  \Lambda\longrightarrow\frac{u}{3}
  \left(\frac{2\pi}{K}\right)^{d-1}.
  \label{eqn_phonon_coupling_expression}
 \end{equation}

  Let us return to the discussion of the coherence length.
  From eqn.~(\ref{eqn_coherence_length}) and using
$\Delta_H\approx\Delta_\mathrm{SC}$, we now obtain
 \begin{equation}
  \xi\approx\frac{u\hbar\sqrt2}{\Delta_\mathrm{SC}}.
 \end{equation}
  That is, the coherence length is inversely correlated with the
SC gap. $u\sim v_F$ is usually not expected to vary drastically.
  From this equation and eqn.~(\ref{eqn_penetration_depth}),
we can eliminate $u$ which is difficult to calculate accurately, and
obtain
 \begin{equation}
  \lambda\xi\Delta_\mathrm{SC}\sqrt{g_F}\approx
  \sqrt{\frac{2\hbar^2}{\mu_0e^2}}.
  \label{eqn_lam_xi_del}
 \end{equation}
  Other than $g_F$, which refers to the normal-state density-of-states,
all parameters on the left-hand side characterize the SC state.
  We have carried out some preliminary estimations based on this
equation, and found that there is indeed an inverse correlation
between $\lambda\xi$ and the SC critical temperature $T_\mathrm{c}$,
both of which are well measured.
  Further work is needed to test eqn.~(\ref{eqn_lam_xi_del}).

 \section{Conclusions}

  We argued that the symmetry-breaking pattern in superconductivity
is given by U(1)$_\mathrm{V}\otimes$U(1)$_\mathrm{A}\to$U(1)$_\mathrm{A}$,
as opposed to
the breaking of U(1)$_\mathrm{V}\equiv$U(1)$_\mathrm{ele/mag}$ by
itself as is commonly thought.

  This allows us to calculate the parameters of the symmetry-breaking
in an essentially non-perturbative way, and we obtained a number of
strikingly simple results which need to be compared with experiment
insofar as this is possible.
  These are:
 \begin{itemize}
%  \item The average charge neutrality of excitations with energy gap $\Delta_\mathrm{SC}$.
  \item The presence of a collective excitation mode (Higgs), which
occurs with a gap that is close to $\Delta_\mathrm{SC}$.
  \item An expression for the BCS phonon coupling, eqn.~(\ref{eqn_phonon_coupling_expression}).
  \item The result that the vacuum expectation value of the SC
condensate is determined almost solely by the normal-state density-of-states,
and simple expressions for the penetration depth and coherence
length. In particular, eqn.~(\ref{eqn_lam_xi_del}) needs to be tested.
 \end{itemize}

  A natural extension of this work will be to the case of the co-existence of
magnetism and SC, especially in the context of high-$T_\mathrm{c}$
superconductivity \cite{highTc}. In this regard, we expect that the
anti-ferromagnetic ratio rule \cite{odagirimagnetism} will
be satisfied in the SC nodal directions, i.e., the density-of-states curve is concave.
  An experimental verification of the same will be a useful initial step
towards the analysis of this problem.

 \section*{Acknowledgement}

  This problem was suggested to the author by T.~Yanagisawa.
  We thank him for discussions and for pointing out a number
of errors in the initial manuscript.

  This work was carried out at Punjabi University, Patiala.
We thank R.~C.~Verma and the supporting staff members of
Punjabi University, Patiala, who made this visit possible.

  This work was supported in part by the University Grants Commission (UGC)
grants for Centre of Advanced Study in Physics, Punjabi University, Patiala.

  We thank I.~Hase, R.~C.~Verma and K.~Yamaji 
for discussions.

 \end{document}